\documentclass{article}
\usepackage{spconf,amsmath,graphicx,hyperref}
\usepackage{multirow}
\usepackage{tabularx}
\usepackage{caption}
\usepackage{arydshln}
\usepackage{booktabs}
\usepackage{xcolor}
\usepackage{graphicx}

\title{Rethinking Fusion: Disentangled Learning of Shared and Modality-Specific Information for Stance Detection}
\name{Zhiyu Xie$^{1,\dag}$, 
Fuqiang Niu$^{2,\dag}$\thanks{$^\dag$ Equal contribution.}, 
Genan Dai$^{1}$, 
Qianlong Wang$^{1}$, 
Li Dong$^{1}$, 
Bowen Zhang$^{1,*}$, 
Hu Huang$^{2,*}$\thanks{$^*$ Corresponding authors}}
\address{$^{1}$School of Artificial Intelligence, Shenzhen Technology University, Shenzhen, China \\
$^{2}$School of Cyber Science and Technology, University of Science and Technology of China, Hefei, China
}
\begin{document}
\ninept
\maketitle
\begin{abstract}
Multi-modal stance detection (MSD) aims to determine an author’s stance toward a given target using both textual and visual content. While recent methods leverage multi-modal fusion and prompt-based learning, most fail to distinguish between modality-specific signals and cross-modal evidence, leading to suboptimal performance. We propose DiME (Disentangled Multi-modal Experts), a novel architecture that explicitly separates stance information into textual-dominant, visual-dominant, and cross-modal shared components. DiME first uses a target-aware Chain-of-Thought prompt to generate reasoning-guided textual input. Then, dual encoders extract modality features, which are processed by three expert modules with specialized loss functions: contrastive learning for modality-specific experts and cosine alignment for shared representation learning. A gating network adaptively fuses expert outputs for final prediction. Experiments on four benchmark datasets show that DiME consistently outperforms strong unimodal and multi-modal baselines under both in-target and zero-shot settings. 
\end{abstract}
\section{Introduction}
\label{sec:intro}

Stance detection identifies an author’s stance (e.g., \textit{favor}, \textit{against} or \textit{neutral}) toward a target and has become an essential task for analyzing online discourse. With the rapid growth of social media, stance detection supports applications such as misinformation detection, public opinion monitoring, and political debate analysis \cite{zhang2025logic, dai2025large}. 
Early research concentrated on text-only models, including feature-based classifiers and neural encoders~\cite{mohammad2016semeval, gomez2023stance}. 
In recent years, user-generated content on social media has become increasingly multi-modal. 
As a result, multi-modal stance detection (MSD) has attracted broad attention. 
MSD aims to infer stance from text–image pairs~\cite{liang2024multimodal}. 
In such posts, text and images co-occur and may convey conflicting cues, which renders MSD both challenging and impactful.

Recent studies commonly adopt dual-stream encoders—e.g., BERT for text and ViT for images—followed by a fusion module \cite{niu2024mmmtcsd}.
However, simple concatenation or late fusion yields limited gains because it fails to capture cross-modal interactions \cite{liang2024multimodal}.
To address this limitation, recent approaches strengthen cross-modal coupling through contrastive cross-modal alignment (e.g., CLIP) and single-stream cross-modal attention (e.g., ViLT) \cite{radford2021clip, kim2021vilt}.
Studies also explore target-aware prompt tuning and adapter-based parameter-efficient tuning to improve semantic integration \cite{liang2024multimodal,houlsby2019adapter, gao2024clipadapter}.
The scope further extends to conversational and zero-shot scenarios, where models must infer stance for unseen targets \cite{allaway2020zero, ding-etal-2025-zero, UPADHYAYA2025104223,Niu2024ACD}.
Moreover, multi-modal large language models have been applied to stance understanding, showing strong potential but raising concerns about bias, stability, and computational efficiency \cite{weinzierl2024tree}.

\begin{figure}
\centering
\includegraphics[width=0.7\linewidth]{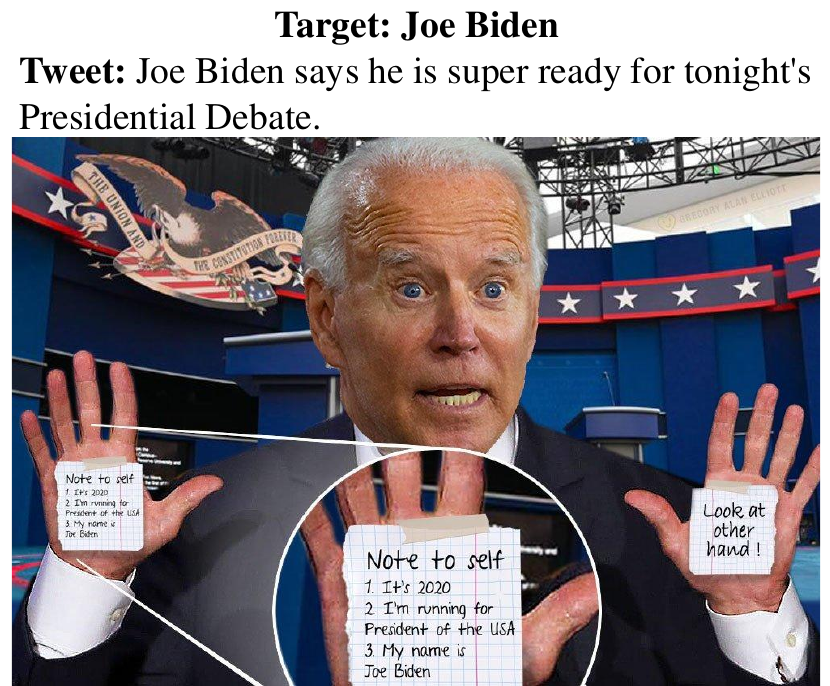}
\caption{An example of a user expressing an ``Against'' stance towards ``Joe Biden'' using multi-modal information.}\vspace{-0.5cm} 
\label{fig:example}
\end{figure}

Despite these advances, two major challenges remain. First, MSD requires deep semantic alignment across heterogeneous modalities: aligning symbolic, context-rich text with visually grounded signals is non-trivial, especially when stance is implicit or sarcastic. Second, stance is often conveyed through fine-grained cross-modal interactions, such as ironic captions paired with symbolic imagery. 
As illustrated in Fig.~\ref{fig:example}, the tweet appears neutral, but the image subtly mocks the subject through visual satire, revealing an “against” stance only when both modalities are interpreted together. 
However, conventional fusion strategies remain coarse and fail to capture such nuances. 
Moreover, few approaches explicitly disentangle shared cross-modal features, which offer reliable stance information, from modality-specific cues, which are often noisy or misleading.

In this work, we introduce \textbf{DiME} (Disentangled Multi-modal Experts), a unified framework that addresses the limitations of shallow fusion by explicitly decomposing stance prediction into three expert pathways: a Textual Expert, a Visual Expert, and an Alignment Expert. Rather than merging modalities uniformly, DiME processes each expert input through a lightweight transformer-based fusion module and applies differentiated supervision to encourage expert specialization. The Textual and Visual Experts learn to focus on modality-dominant features via triplet-based contrastive objectives, while the Alignment Expert captures shared, cross-modal information through cosine-based consistency losses. Their outputs are dynamically weighted using a gating network conditioned on the input features, and the fused result is passed to a final classification layer. This design enables DiME to handle fine-grained cross-modal interactions and suppress modality-specific noise, resulting in more robust stance predictions. 

Our contributions are threefold:  
(i) We propose DiME, a Multi-Expert architecture that explicitly disentangles modality-specific and cross-modal features via dedicated Textual, Visual, and Alignment experts, overcoming the limitations of shallow fusion.
(ii) We design a differentiated objective scheme that guides each expert using tailored loss functions, enabling specialization and reducing cross-modal noise during training.
(iii) We conduct comprehensive experiments across four benchmarks and multiple settings (in-target, zero-shot, and ablation), demonstrating that DiME achieves state-of-the-art performance and strong generalization.

\section{Methodology}
\label{sec:METHODOLOGY}

\begin{figure}
\centering
\includegraphics[width=0.9\linewidth]{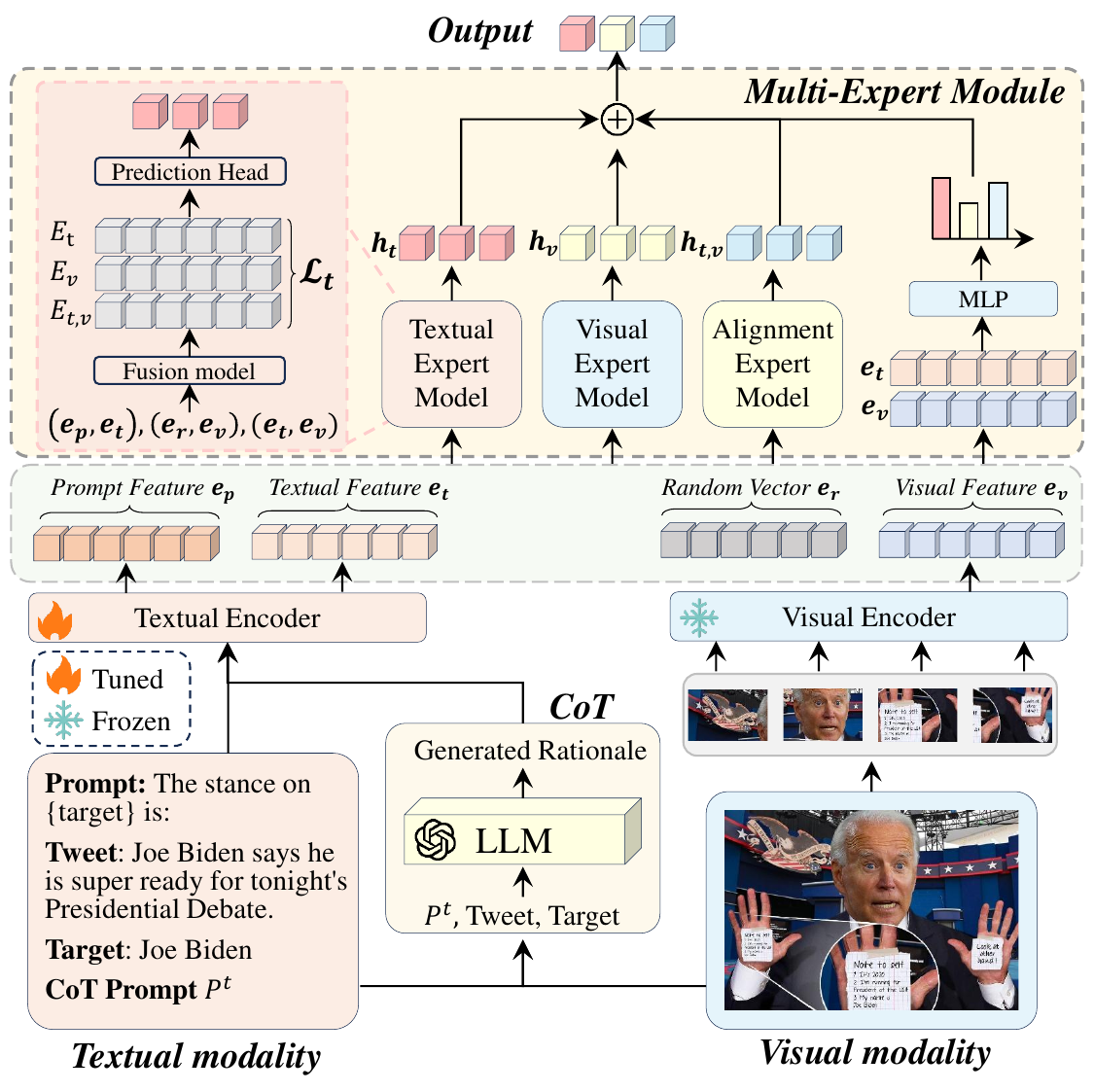}
\caption{The overall architecture of our proposed DiME.} \vspace{-0.5cm}
\label{fig:model}
\end{figure}

\vspace{-0.2cm}
\subsection{Task Definition}
In multi-modal stance detection, each instance consists of text $S$, image $I$, and an explicit target $t$; the goal is to assign a stance label $y\in\{{Favor}, {Against}, {Neutral}\}$.

\vspace{-0.3cm}
\subsection{Method Overview}

As illustrated in Fig.~\ref{fig:model}, DiME~\footnote{\url{https://github.com/PhilomathTse/DiME}} consists of three components: (i) a target-aware Chain-of-Thought prompt that elicits an LLM rationale to expose implicit text–image relations and augment the input text; (ii) textual and visual encoders that read the target+post+rationale and the image with a lightweight random visual prompt to obtain compact representations; and (iii) a multi-expert module with the Textual, Visual, and Alignment Experts to model modality-specific and shared factors. A lightweight gating network computes instance-wise mixture weights from both modalities to fuse expert outputs, and a classification head produces the final stance.

\subsection{Reasoning-Chain Generation}
Given a multi-modal post $(S,I)$ and target $t$, we first obtain a textual \emph{reasoning chain} $R$ that verbally explains how the text and image together express a stance toward $t$. We use a multi-modal LLM  and query it with a structured CoT prompt $P^t$ to elicit an explicit analysis. 
\begin{center} \fcolorbox{black}{blue!5}{\parbox{0.95\linewidth}{$P^t$: {Give you a sentence, an image and target, please analyze what stance the sentence and the image express toward the target. The text you need to analyze is: ``\%s'' about target ``\%s''. Your output: Analysis: Your analysis of the stance.}}} 
\end{center}
\vspace{-0.2cm}

\subsection{Encoders}
We encode text and image with separate encoders and project all features to a common d-dimensional space.

\textbf{Textual Encoder.}
We adopt a BERT-based encoder to obtain stance-aware textual features. The input consists of: (i) a short task prompt $P^{m}$, (ii) the post text $S$, and (iii) the CoT rationale $R$. We first encode $P^{m}$ alone to obtain a prompt feature $e_{p}$. Then we concatenate $S$ and $R$ with a \texttt{[SEP]} token and feed them into the same encoder to obtain a content feature $e_{t}$. Both $e_{p}$ and $e_{t}$ are linearly projected to dimension $d$ and used as the textual representation.

\textbf{Visual Encoder.}
Following ViT~\cite{dosovitskiy2021an}, we split the image $I$ into non-overlapping patches, linearly embed them, add positional encodings, and pass the sequence through a transformer. We take the \texttt{[CLS]} token as the visual feature $e_{v}$ and project it to dimension $d$ with a linear layer.

\vspace{-0.2cm}
\subsection{Multi-Expert Module}
We employ three experts with identical pipelines but distinct learning objectives. Given the encoder outputs, we construct a feature set
$\{(e_p, e_t),\ (e_r, e_v),\ (e_t, e_v)\}$,
where $e_p$ is the prompt feature, $e_t$ the textual feature, $e_v$ the visual feature, and $e_r\!\in\!{R}^{d}$ a dimension-$d$ random vector. The random vector acts as a lightweight stochastic visual prompt and feature-space regularizer, enriching visual semantics and mitigating co-adaptation. Each pair is fused by a lightweight Transformer block (denoted Fuse), yielding three fused representations:
\[
E_t=\mathrm{Fuse}(e_p,e_t),\quad
E_v=\mathrm{Fuse}(e_r,e_v),\quad
E_{t,v}=\mathrm{Fuse}(e_t,e_v).
\]
The experts operate on these fused representations with distinct objectives.

\textbf{Textual Expert.}
The Textual Expert aims to capture textual-dominant cues and suppress spurious visual interference. We construct a triplet $(E_t, E_v, E_{t,v})$ with anchor $E_{t,v}$, positive $E_t$, and negative $E_v$, and optimize a triplet margin loss \cite{Schroff_2015_CVPR} with Euclidean distance $d(\cdot,\cdot)$ :
\[
\mathcal{L}_{T}=\max\!\big(0,\ m + d(E_{t,v},E_t) - d(E_{t,v},E_v)\big),
\]
where $m$ is a margin hyperparameter.

\textbf{Visual Expert.}
The Visual Expert mirrors the Textual Expert, encouraging sensitivity to visual-dominant cues while being robust to textual distractions. Using the triplet $(E_v, E_t, E_{t,v})$, we optimize
\[
\mathcal{L}_{V}=\max\!\big(0,\ m + d(E_{t,v},E_v) - d(E_{t,v},E_t)\big).
\]

\textbf{Alignment Expert.}
The Alignment Expert targets cross-modal common information that is consistent across modalities. We supervise it with cosine-consistency to pull the cross-modal fusion toward both unimodal fusions:
\[
\mathcal{L}_{S}= \big(1-\cos(E_{t,v},E_t)\big) + \big(1-\cos(E_{t,v},E_v)\big).
\]

\textbf{Fusion and Prediction.}
Each expert outputs a modality-specific representation: $h_t$, $h_v$, and $h_{t,v}$. To combine them, we use a gating network that generates adaptive weights based on the input features. Specifically, we concatenate the encoder outputs $(e_t, e_v)$ and feed them into a two-layer MLP followed by a temperature-scaled softmax to obtain the fusion weights $\boldsymbol{\pi} = [\pi_t, \pi_v, \pi_{t,v}]$:

\[
\boldsymbol{\pi} = \mathrm{softmax}\left( \frac{W_2\, \mathrm{ReLU}(W_1[e_t; e_v] + b_1) + b_2}{\tau} \right)
\]
Here, $W_1, W_2$ and $b_1, b_2$ are gating parameters, and $\tau$ is a temperature hyperparameter controlling the sharpness of the distribution. The final fused representation $h$ is computed as a weighted sum of the expert outputs:
\[
h = \pi_t \cdot h_t + \pi_{t,v} \cdot h_{t,v} + \pi_v \cdot h_v
\]
This fused vector is passed to a softmax classifier to predict the stance label:

\[
\hat{y} = \mathrm{softmax}(W_c h + b_c)
\]
where $W_c$ and $b_c$ are parameters of the final classification layer. This fusion mechanism allows the model to dynamically adjust the importance of each expert based on input semantics, enabling more precise stance prediction.

\vspace{-0.3cm}
\subsection{Training Objective}
We combine the expert objectives with the task-level cross-entropy:
\[
\mathcal{L}_{CE} = -\sum_{c\in\mathcal{Y}} y_c\log \hat{y}_c, \qquad
\mathcal{L}=\mathcal{L}_T+\mathcal{L}_V+\mathcal{L}_S+\mathcal{L}_{CE}.\vspace{-0.2cm}
\]

This objective encourages the Textual and Visual Experts to specialize on their respective modalities while the alignment expert consolidates modality-invariant evidence, with the gating mechanism aggregating their predictions into a robust final decision.

\vspace{-0.2cm}
\section{Experiment}
\label{sec:EXPERIMENT}

\subsection{Datasets and evaluation metrics}
We evaluate the effectiveness of DiME on the widely used multi-modal stance detection datasets in the MMSD benchmark \cite{liang2024multimodal}, including MTSE, MCCQ, MWTWT and MRUC.
Following MMSD, Each dataset is divided into training, development, and testing sets with a ratio of 7:1:2.
Following \cite{liang2024multimodal}, we adopt the macro-F1 as the primary evaluation metric to quantify the performance of the multi-modal stance model.

\vspace{-0.3cm}
\subsection{Implementation details}
We use a lightweight Transformer encoder as the multi-modal fusion module. Text features (768-d from BERT-base-uncased) are projected to 512-d and normalized, while visual features (512-d from CLIP ViT-B/32) are similarly normalized. The two embeddings are treated as tokens and passed into a Transformer encoder (d\_model=256, 4 heads, 1–2 layers, FFN dimension=512, GELU activation, dropout=0.1). Mean pooling over the output tokens yields a fused representation, which is fed into a linear classification head. The multi-objective losses decrease smoothly and typically converge within 15 epochs.
\vspace{-0.3cm}
\subsection{Compared Baseline Methods}
We compare with unimodal and multi-modal baselines. Textual models include BERT \cite{devlin2019bert}, RoBERTa \cite{liu2019roberta}, KEBERT \cite{kawintiranon-singh-2021-knowledge}, LLaMA2 \cite{touvron2023llama}, and GPT-4. Visual models include ResNet \cite{he2016deep}, ViT \cite{dosovitskiy2021an}, and Swin Transformer \cite{liu2021swin}. Multi-modal baselines include ViLT \cite{kim2021vilt}, CLIP \cite{radford2021clip}, BERT+ViT, Qwen-VL \cite{bai2024qwenvl}, GPT-4o, and TMPT/TMPT + CoT \cite{liang2024multimodal}.

\section{Experimental Results}
\label{sec:EXPERIMENT}

\subsection{Main Results}
Our experimental results on four MMSD benchmark datasets (MTSE, MCCQ, MWTWT, and MRUC) demonstrate that our proposed DiME model consistently outperforms a wide range of strong unimodal and multi-modal baselines, verifying its effectiveness in MSD.
As shown in Table~\ref{tab:in_target}, DiME achieves the highest overall macro-F1 average (68.95), with statistically significant improvements ($p$-value $<$ 0.05) over previous models, confirming the robustness of our architecture.

Specifically, we observe that while visual-only methods perform poorly across all datasets due to their limited stance expressiveness, text-only methods—especially prompt-based or knowledge-enhanced ones such as KEBERT and GPT-4o—achieve moderate results. However, they still fall short of fully leveraging the multi-modal signals. Among existing multi-modal baselines, pre-trained visual-language models such as CLIP and ViLT outperform early fusion models (e.g., BERT+ViT), and prompt-tuning approaches (e.g., TMPT and TMPT+CoT) further improve performance through lightweight adaptation. Despite this, DiME surpasses all these methods with large margins, including improvements of +2.90 over KEBERT (66.05), +4.65 over CLIP (64.30), and +1.83 over TMPT+CoT (67.12).

In particular, DiME achieves state-of-the-art performance on 8 out of 10 target domains. Notably, the model reaches 88.04 on DF, 65.76 on UKR, and 70.85 on JB—three of the most visually and semantically challenging settings—demonstrating its strong capacity for extracting robust cross-modal stance signals. On CE and AH, where stance cues are more explicit and less reliant on modality interaction, other methods such as TMPT+CoT yield similar or slightly higher results. This suggests that in highly textual-dominant cases, the added regularization from the alignment expert may be less beneficial and could even constrain performance, pointing to the potential for adaptive expert gating in future work.

These findings suggest that DiME’s core design—explicit disentanglement of modality-specific and shared information, paired with expert specialization and gated fusion—is particularly beneficial for handling sarcasm, indirect expression, or visual symbolism, which are common in real-world political and social media content. Furthermore, DiME demonstrates strong generalization across both rich multi-modal and textual-dominant scenarios, making it a highly versatile solution for stance detection.

\begin{table*}
\small
\centering
\caption{Experimental results (\%) of in-target multi-modal stance detection. The best scores in each group are highlighted in bold. The second best value is underlined. Results marked with $^\star$ indicate significance at p-value $<$ 0.05.}

\label{tab:in_target}
\resizebox{\linewidth}{!}{
\begin{tabular}{llcclclccccclccll}
\hline
\multirow{2}{*}{\textbf{Modality}} 
  & \multirow{2}{*}{\textbf{Method}} 
  & \multicolumn{2}{c}{\textbf{MTSE}} & 
  &\multicolumn{1}{c}{\textbf{MCCQ}} &
  &\multicolumn{5}{c}{\textbf{MWTWT}} & 
  & \multicolumn{2}{c}{\textbf{MRUC}} & 
    &\multirow{2}{*}{\textbf{Avg.}}\\
\cline{3-4} \cline{6-6} \cline{8-12} \cline{14-15}
& & DT & JB  && CQ  && CA & CE & AC & AH &DF && RUS & UKR  & & \\
\hline
\multirow{5}{*}{Textual}
 & BERT     & 48.25 & 52.04  && 66.57  && 75.62 & 60.85 & 63.05 & 59.24 & 81.53  && 41.25 & 46.80  & &59.52\\
 & RoBERTa  & 58.39 & 60.79  && 66.57  && 69.56 & 65.03 & \underline{69.74}& 67.99 & 79.21  && 39.52 & 57.66  & &63.45\\
 & KEBERT   & 64.50 & \underline{69.81}&& 66.84  && 71.67 & \underline{67.56}& 69.29 & \underline{69.74}& 80.57  && 41.55 & 59.01  & &66.05\\
 & LLaMA2   & 53.23 & 52.67  && 47.40  && 34.89 & 41.95 & 49.09 & 44.32 & 30.21  && 38.84 & 38.54  & &43.11\\
 & GPT-4o$^\star$     & 63.97 & 69.69  && 63.08  && 44.87 & 46.57 & 43.96 & 44.64 & 57.17  && 48.97 & 51.53  & &53.45 \\
\hline
\multirow{3}{*}{Visual}
 & ResNet   & 37.89 & 38.59  && 47.16  && 39.89 & 42.20 & 43.52 & 37.05 & 50.34  && 35.10 & 40.00  & &41.17\\
 & ViT      & 40.48 & 40.42  && 46.64  && 46.63 & 50.00 & 40.16 & 46.32 & 50.86  && 33.31 & 39.87  & &43.47\\
 & SwinT    & 39.89 & 40.43  && 48.80  && 46.30 & 46.99 & 41.02 & 47.39 & 51.32  && 35.01 & 40.89  & &43.80\\
\hline
\multirow{8}{*}{Multi-modal}
 & BERT+ViT    & 41.86 & 45.82  && 61.32  && 63.20 & 44.71 & 56.45 & 46.85 & 73.71  && 39.28 & 48.41  & &52.16\\
 & ViLT        & 35.32 & 48.24  && 47.85  && 62.70 & 56.44 & 58.06 & 60.22 & 73.66  && 34.62 & 42.41  & &51.95\\
 & CLIP        & 53.22 & 65.83  && 63.65  && 70.93 & 67.17 & 67.43 & \textbf{70.86}& 79.06  && 44.99 & 59.86  & &64.30\\
 & Qwen-VL     & 43.31 & 45.13  && 50.51  && 43.06 & 45.49 & 49.79 & 46.04 & 27.73  && 36.50 & 40.78  & &42.83\\
 & GPT-4o$^\star$  & 48.90  & 43.41  && 48.83  && 52.96  & 61.47  & 49.63  & 53.41  & 58.67  && \underline{45.53}& 51.32  && 51.41\\
 & TMPT        & 55.41 & 61.61  && 67.67  && \underline{76.60}& 63.19 & 67.25 & 62.92 & 81.19  && 43.56 & 59.24  & &63.86\\
 & TMPT+CoT    & \underline{66.61} & 68.75  && \underline{71.79}&& 74.40 & \textbf{69.96} & 68.43 & 63.00 & \underline{82.71}&& 45.04 & \underline{60.52}& &\underline{67.12}\\
  \cdashline{2-17}
 &\textbf{DiME$^\star$}   & \textbf{67.81} & \textbf{70.85}  && \textbf{72.46}  && \textbf{77.26} & 65.50 & \textbf{70.31} & 65.51& \textbf{88.04}  && \textbf{45.98} & \textbf{65.76}  & &\textbf{68.95}\\
\hline
\end{tabular}
}

\end{table*}

\subsection{Zero-shot Results}

We further assess the generalization capability of DiME in a zero-shot setting, where stance targets in the test set are entirely unseen during training. The results across MTSE, MWTWT, and MRUC are summarized in Table~\ref{tab:zero-shot}, with macro-F1 as the evaluation metric.

DiME achieves the highest performance on MTSE (60.29) and MRUC (53.83), outperforming all competing models across unimodal and multi-modal settings. On MWTWT, DiME reaches 64.31, which is slightly lower than TMPT (65.16), yet still among the top-performing methods. Compared to strong multi-modal baselines such as CLIP and Qwen-VL, DiME shows notable improvements, particularly in MTSE (+3.91 over TMPT+CoT and +13.41 over Qwen-VL), reflecting its robustness when transferring to unseen targets.

These results confirm that DiME’s design—especially its expert disentanglement and CoT-enhanced textual reasoning—facilitates stronger generalization across targets. Even in the absence of target-specific supervision, DiME effectively leverages shared and modality-specific cues, making it better suited for real-world scenarios where new stance targets emerge dynamically.

\begin{table}
\centering
\small
\caption{The reported results are the macro-F1 across all targets in a dataset on zero-shot multi-modal stance detection.}
\label{tab:zero-shot}
\begin{tabular}{llccc}
\hline
\textbf{Modality}
  & {\textbf{Method}} 
  & {\textbf{MTSE}} 
  &{\textbf{MWTWT}} 
  & {\textbf{MRUC}} 

\\

\hline
\multirow{5}{*}{Textual}
 & BERT     &31.25 &59.23 &18.73   \\
 & RoBERTa  &29.41 &60.19 &23.54   \\
 & KEBERT   &28.99 &60.43 &26.43   \\
 & LLaMA2   &53.75 &41.26 &34.10   \\
\midrule
\multirow{3}{*}{Visual}
 & ResNet   &27.61 &24.40 &24.73   \\
 & ViT      &29.17 &30.06 &27.89  \\
 & SwinT    &29.70 &31.81 &24.99   \\
\midrule
\multirow{7}{*}{Multi-modal}
 & BERT+ViT    &29.14 &60.71 &19.27  \\
 & ViLT        &28.91 &46.97 &22.76  \\
 & CLIP        &28.60 &60.15 &26.51  \\
 & Qwen-VL     &46.88 &42.69 &39.17  \\
 & TMPT        &32.17 &\textbf{65.16} &24.29 \\
 & TMPT+CoT    &\underline{56.38} &62.56 &\underline{50.37}  \\
 & \textbf{DiME}        &\textbf{60.29} &\underline{64.31} &\textbf{53.83}  \\
\hline
\end{tabular}

\end{table}

\begin{figure}
\centering
\includegraphics[width=\linewidth]{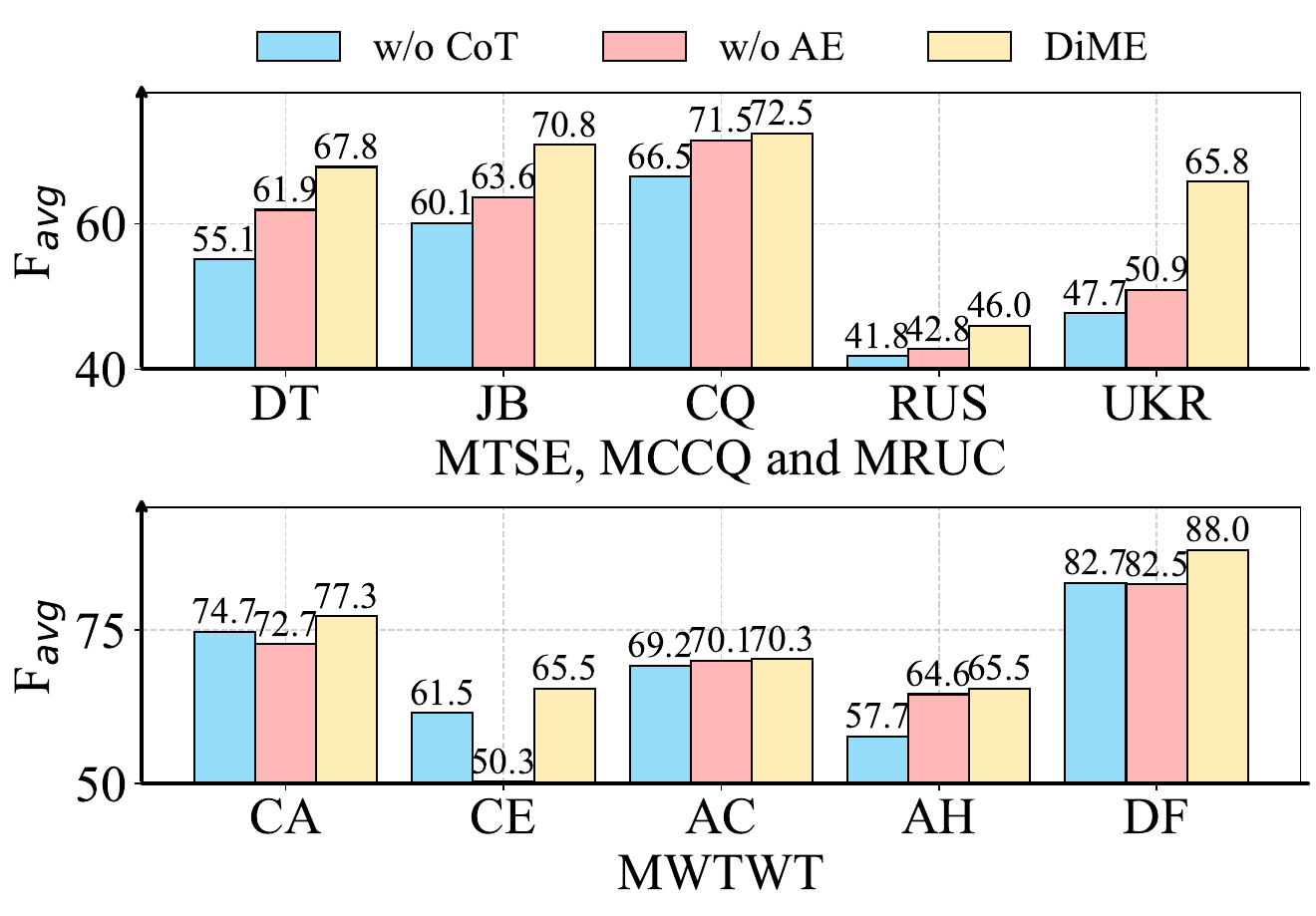}
\caption{Experimental results of ablation study.} \vspace{-0.2cm}
\label{ablation}
\end{figure}

\subsection{Ablation Study}

We evaluate the contribution of two core components in DiME by ablating the reasoning-chain (w/o CoT) generation and the Alignment Expert (w/o AE), keeping all settings identical to the main experiments; results over ten targets are shown in Fig.\ref{ablation}. Removing CoT consistently degrades performance, with the largest drops on textual-dominant or reasoning-sensitive targets such as DT, CE, and AH, where macro-F1 decreases by more than five points. This suggests that CoT rationales enrich the textual input with target-aware reasoning, which is especially helpful when stance cues are implicit or subtle. Likewise, removing the Alignment Expert notably harms targets that require multi-modal coordination, including UKR, DF, and RUS; for instance, on DF, DiME attains 88.0, whereas the w/o AE variant drops to 82.5. These trends underscore the value of explicitly modeling shared cross-modal factors to capture text–image alignment signals. Overall, both components contribute positively and complementarily: CoT strengthens implicit and logic-heavy textual reasoning, while AE enhances cross-modal alignment under visually contrastive or politically sensitive content. Their combination yields robust gains across diverse stance expressions, including sarcasm, implicit reasoning, and cross-modal inconsistency.

\section{CONCLUSION}
\label{sec:CONCLUSION}
We proposed DiME, a multi-expert framework for MSD that explicitly disentangles textual-specific, visual-specific, and shared cross-modal information. The architecture integrates reasoning-aware textual inputs, dual encoders, and expert modules trained with differentiated objectives. DiME achieves state-of-the-art results across four widely used MSD benchmarks and demonstrates strong performance in both in-target and zero-shot evaluation. Our results highlight the value of separating modality roles and tailoring supervision accordingly. Future work will explore dynamic expert interaction and applications to broader multi-modal reasoning tasks.

\vfill\pagebreak
\section{Acknowledgments}

This research was supported by the Key-Area Research and Development Program of Guangdong Province (No. 2025B0101130002); the National Natural Science Foundation of China (No. 62306184); the Natural Science Foundation for Top Talents of SZTU (No. GDRC202518 and No. GDRC202320); the Shenzhen Science and Technology Program (No. RCBS20231211090548077 and No. JCYJ20240813113218025); and the Project for Improving Scientific Research Capabilities of Key Construction Disciplines in Guangdong Province (No. 2025ZDJS039).

\bibliographystyle{IEEEbib}
\bibliography{refs}

\begin{thebibliography}{10}

\bibitem{zhang2025logic}
Bowen Zhang, Jun Ma, Xianghua Fu, and Genan Dai,
\newblock ``Logic augmented multi-decision fusion framework for stance detection on social media,''
\newblock {\em Information Fusion}, p. 103214, 2025.

\bibitem{dai2025large}
Genan Dai, Jiayu Liao, Sicheng Zhao, Xianghua Fu, Xiaojiang Peng, Hu~Huang, and Bowen Zhang,
\newblock ``Large language model enhanced logic tensor network for stance detection,''
\newblock {\em Neural Networks}, vol. 183, pp. 106956, 2025.

\bibitem{mohammad2016semeval}
Saif Mohammad, Svetlana Kiritchenko, Parinaz Sobhani, Xiaodan Zhu, and Colin Cherry,
\newblock ``Semeval-2016 task 6: Detecting stance in tweets,''
\newblock in {\em Proceedings of SemEval-2016}, 2016, pp. 31--41.

\bibitem{gomez2023stance}
Manuela G{\'o}mez-Suta, Juli{\'a}n Echeverry-Correa, and Jos{\'e}~A Soto-Mej{\'\i}a,
\newblock ``Stance detection in tweets: A topic modeling approach supporting explainability,''
\newblock {\em Expert Systems with Applications}, vol. 214, pp. 119046, 2023.

\bibitem{liang2024multimodal}
Bin Liang, Ang Li, Jingqian Zhao, Lin Gui, Min Yang, Yue Yu, Kam{-}Fai Wong, and Ruifeng Xu,
\newblock ``Multi-modal stance detection: New datasets and model,''
\newblock in {\em Findings of ACL 2024}, 2024, pp. 12373--12387.

\bibitem{niu2024mmmtcsd}
Fuqiang Niu, Zebang Cheng, Xianghua Fu, Xiaojiang Peng, Genan Dai, Yin Chen, Hu~Huang, and Bowen Zhang,
\newblock ``Multimodal multi-turn conversation stance detection: A challenge dataset and effective model,''
\newblock in {\em Proceedings of ACM MM 2024}, 2024, pp. 3867--3876.

\bibitem{radford2021clip}
Alec Radford, Jong~Wook Kim, Chris Hallacy, Aditya Ramesh, Gabriel Goh, Sandhini Agarwal, Girish Sastry, Amanda Askell, Pamela Mishkin, Jack Clark, Gretchen Krueger, and Ilya Sutskever,
\newblock ``Learning transferable visual models from natural language supervision,''
\newblock in {\em ICML 2021}, 2021, pp. 8748--8763.

\bibitem{kim2021vilt}
Wonjae Kim, Bokyung Son, and Ildoo Kim,
\newblock ``{ViLT}: Vision-and-language transformer without convolution or region supervision,''
\newblock in {\em ICML 2021}, 2021, pp. 5583--5594.

\bibitem{houlsby2019adapter}
Neil Houlsby, Andrei Giurgiu, Stanislaw Jastrzebski, Bruna Morrone, Quentin de~Laroussilhe, Andrea Gesmundo, Mona Attariyan, and Sylvain Gelly,
\newblock ``Parameter-efficient transfer learning for {NLP},''
\newblock in {\em ICML 2019}, 2019, pp. 2790--2799.

\bibitem{gao2024clipadapter}
Peng Gao, Shijie Geng, Renrui Zhang, Teli Ma, Rongyao Fang, Yongfeng Zhang, Hongsheng Li, and Yu~Qiao,
\newblock ``{CLIP}-adapter: Better vision-language models with feature adapters,''
\newblock {\em International Journal of Computer Vision}, vol. 132, no. 2, pp. 581--595, 2024.

\bibitem{allaway2020zero}
Emily Allaway and Kathleen McKeown,
\newblock ``Zero-shot stance detection: A dataset and model using generalized topic representations,'' 2020.

\bibitem{ding-etal-2025-zero}
Yuzhe Ding, Kang He, Bobo Li, Li~Zheng, Haijun He, Fei Li, Chong Teng, and Donghong Ji,
\newblock ``Zero-shot conversational stance detection: Dataset and approaches,''
\newblock in {\em Findings of the Association for Computational Linguistics: ACL 2025}, pp. 3221--3235.

\bibitem{UPADHYAYA2025104223}
Apoorva Upadhyaya, Wolfgang Nejdl, and Marco Fisichella,
\newblock ``Interpretable zero-shot stance detection with proactive content intervention,''
\newblock {\em Information Processing \& Management}, p. 104223, 2025.

\bibitem{Niu2024ACD}
Fuqiang Niu, Min Yang, Ang Li, Baoquan Zhang, Xiaojiang Peng, and Bowen Zhang,
\newblock ``A challenge dataset and effective models for conversational stance detection,''
\newblock in {\em International Conference on Language Resources and Evaluation}, 2024.

\bibitem{weinzierl2024tree}
Maxwell Weinzierl and Sanda Harabagiu,
\newblock ``Tree-of-counterfactual prompting for zero-shot stance detection,''
\newblock in {\em Proceedings of ACL 2024}, 2024.

\bibitem{dosovitskiy2021an}
Alexey Dosovitskiy, Lucas Beyer, Alexander Kolesnikov, Dirk Weissenborn, Xiaohua Zhai, Thomas Unterthiner, Mostafa Dehghani, Matthias Minderer, Georg Heigold, Sylvain Gelly, Jakob Uszkoreit, and Neil Houlsby,
\newblock ``An image is worth 16x16 words: Transformers for image recognition at scale,''
\newblock in {\em International Conference on Learning Representations}, 2021.

\bibitem{Schroff_2015_CVPR}
Florian Schroff, Dmitry Kalenichenko, and James Philbin,
\newblock ``Facenet: A unified embedding for face recognition and clustering,''
\newblock in {\em Proceedings of the IEEE Conference on Computer Vision and Pattern Recognition (CVPR)}, June 2015.

\bibitem{devlin2019bert}
Jacob Devlin, Ming{-}Wei Chang, Kenton Lee, and Kristina Toutanova,
\newblock ``Bert: Pre-training of deep bidirectional transformers for language understanding,''
\newblock in {\em NAACL-HLT 2019}, 2019, pp. 4171--4186.

\bibitem{liu2019roberta}
Yinhan Liu, Myle Ott, Naman Goyal, Jingfei Du, Mandar Joshi, Danqi Chen, Omer Levy, Mike Lewis, Luke Zettlemoyer, and Veselin Stoyanov,
\newblock ``Roberta: A robustly optimized {BERT} pretraining approach,'' 2019.

\bibitem{kawintiranon-singh-2021-knowledge}
Kornraphop Kawintiranon and Lisa Singh,
\newblock ``Knowledge enhanced masked language model for stance detection,''
\newblock in {\em Proceedings of the 2021 Conference of the North American Chapter of the Association for Computational Linguistics: Human Language Technologies}. 2021, Association for Computational Linguistics.

\bibitem{touvron2023llama}
Hugo Touvron, Louis Martin, Kevin Stone, Peter Albert, Amjad Almahairi, Yasmine Babaei, Nikolay Bashlykov, Soumya Batra, Prajjwal Bhargava, Shruti Bhosale, et~al.,
\newblock ``Llama 2: Open foundation and fine-tuned chat models,''
\newblock {\em arXiv preprint arXiv:2307.09288}, 2023.

\bibitem{he2016deep}
Kaiming He, Xiangyu Zhang, Shaoqing Ren, and Jian Sun,
\newblock ``Deep residual learning for image recognition,''
\newblock in {\em Proceedings of the IEEE conference on computer vision and pattern recognition}, 2016, pp. 770--778.

\bibitem{liu2021swin}
Ze~Liu, Yutong Lin, Yue Cao, Han Hu, Yixuan Wei, Zheng Zhang, Stephen Lin, and Baining Guo,
\newblock ``Swin transformer: Hierarchical vision transformer using shifted windows,''
\newblock in {\em Proceedings of the IEEE/CVF international conference on computer vision}, 2021, pp. 10012--10022.

\bibitem{bai2024qwenvl}
Jinze Bai, Shuai Bai, Shusheng Yang, Shijie Wang, Sinan Tan, Peng Wang, Junyang Lin, Chang Zhou, and Jingren Zhou,
\newblock ``Qwen-{VL}: A versatile vision-language model for understanding, localization, text reading, and beyond,'' 2024.

\end{thebibliography}

\end{document}